\documentclass[pra,superscriptaddress,showpacs,twocolumn]{revtex4-1}

\usepackage{amsmath,amssymb}
\usepackage{amsfonts}
\usepackage{graphics}
\usepackage{graphicx}
\usepackage{verbatim}
\usepackage{longtable}
\usepackage{mathtools}
\usepackage{float}

\newcommand{\beq}{\begin{eqnarray}}
\newcommand{\eeq}{\end{eqnarray}}

\begin{document}

\title{Collective excitations of fractional quantum Hall states
in monolayer graphene}

\author{Sujit Narayanan and Malcolm P. Kennett}

\affiliation{Department of Physics, Simon Fraser University, 8888 University Drive, Burnaby, British Columbia, V5A 1S6, Canada}

\date{\today}

\begin{abstract}
We study the collective excitations of fractional quantum Hall states in graphene. We focus on states which allow for chiral symmetry breaking (CSB) orders, specifically antiferromagnetism and charge density wave order.  We investigate numerically how the collective excitation spectra depend on filling and the flux attachment scheme for two classes of variational states, the T\"{o}ke-Jain sequence and the Modak-Mandal-Sengupta sequence.
\end{abstract}

\maketitle

\section{Introduction}
Low energy electrons in graphene display a linear dispersion and can be described by a Dirac Hamiltonian \cite{Novoselov2005, Goerbig-review}. As a two dimensional electron gas, when graphene is placed in strong perpendicular magnetic field, it can exhibit the integer quantum Hall effect (IQHE). The pseudo-relativistic nature of electrons in graphene leads to an IQHE which differs from non-relativistic systems. Specifically the energies of Landau levels in a non-relativistic system scale with the Landau level index $n$, while in the relativistic case the energies scale as $\sqrt{n}$. Additionally, the four-fold degeneracy of the zeroth Landau level in graphene implies a multi-component IQHE.\par
The Hall plateaux for non-interacting electrons in graphene are at fillings $\nu = \pm 2 (2n +1) $ for integer $n$ \cite{Novoselov2005, Goerbig-review, Zhang2005} as opposed to $\nu = 2n$ for spin-degenerate non-relativistic systems. Electron-electron interactions in graphene can lead to IQH states at additional filling fractions including $\nu=0, \pm 1$ \cite{Skachko2010,Abanin2013,Yu2013,Young2012,Du2009,Dean2011,Feldman2012,Roy2014,Herbut2007a,Herbut2007b,Herbut2008,Semenoff2011,BR-BLG,Barlas2012,Kharitonov2012, Khveshchenko2001, Leal2004, Herbut2006, HJR2009, Yang2007}. The states at $\nu=0, \pm 1$ cannot be accounted for in a picture of non-interacting electrons \cite{Roy2014} and various proposals have been put forward to explain their origin either in terms of chiral symmetry breaking (CSB) orders such as charge density wave (C) and antiferromagnetism (N) \cite{Khveshchenko2001, Herbut2006, HJR2009, Herbut2007a, Herbut2007b} or in terms of valley-odd quantum Hall ferromagnetism (QHFM) \cite{Semenoff2011, Barlas2012, Yang2007}. 
In the CSB scenario interactions lead to CSB orders that break sublattice symmetry and this has also been demonstrated to give good agreement with experimental results \cite{Roy2014, Hank2020}. \par
The fractional quantum Hall effect (FQHE) has also been observed in graphene \cite{Du2009,Skachko2010,Dean2011,Feldman2012,Feldman2013,Bolotin2009,Amet2015} 
and some experiments have revealed an unusual pattern of fractions that follows the standard composite fermion sequence between filling factors $\nu = 0$ and $\nu = 1$ but involves only even-numerator fractions between $\nu = 1$ and $\nu =2$ \cite{Feldman2012}. Theoretically, the FQHE in graphene has attracted considerable interest \cite{Balram2015, Csaba2007,Csaba2006,Modak2011, Frassdorf2018, Nayak2014, Goerbig2008, Goerbig2007, Papic2010, Papic2009, Scarola2001, Smith2010, Khveshchenko2007, Yang2006, Sodemann2014, Nomura2009, Hegde2022, Dora2022}. 
In particular, Refs. \cite{Modak2011,Khveshchenko2007,Narayanan2018,Cai2013} used the framework of the Chern-Simons theory of multicomponent FQH states in graphene in 
the presence of symmetry breaking orders to investigate possible composite fermion wave functions. Using this framework and allowing for CSB orders, we proposed variational wavefunctions \cite{Narayanan2018} to explain the recent observation of even denominator fractional quantum Hall (EDFQH) states for filling fractions $\nu=\frac{1}{2}$ and $\frac{1}{4}$   \cite{Zibrov2018,Zhou2021}. It has not yet been determined whether these CSB orders are present in experiments. The collective excitations of  FQHE states in graphene may help to distinguish between different possible symmetry breaking orders.

Lopez and Fradkin showed how a system of electrons coupled to a Chern-Simons gauge field is equivalent to a system of composite fermions \cite{Lopez1991}. Employing a random phase approximation method they were able to arrive at expressions for the electromagnetic response of these states for finite wavevector, $\vec{q}$, and frequency, $\omega$, \cite{Chen}. Halperin, Lee and Read \cite{HLR} extended their methods to study the FQH state at $\nu=\frac{1}{2}$ and Simon and Halperin \cite{Simon-Halperin} further developed this approach by taking into account the corrections due to mass renormalization that the Chern-Simons term induces. They also studied the collective excitations of the FQH liquid. These results are consistent with experiments involving using inelastic light scattering \cite{Pinczuk1993} and observation of geometric resonances in the cyclotron orbits of composite fermions (CFs) using surface acoustic waves \cite{Willett1993}. 

Recently the methods developed by Fradkin and Lopez \cite{Lopez1993} have been applied to the case of graphene and expressions for the components of the electromagnetic response tensor were obtained \cite{Frassdorf2018}. However these calculations did not take into account any form of symmetry breaking orders, originally emphasized in Ref. \cite{Kveshchenko}.

In this work we are primarily interested in studying the collective excitations of FQH states for which the filling fraction lies between $0 < \nu <1$. For calculational convenience we focus on the following symmetry breaking orders: out of plane antiferromagnetism, charge density wave and ferromagnetism, since these are easily accommodated in the Chern-Simons theory we employ \cite{Modak2011, Narayanan2018}. We work in the zeroth Landau level (ZLL)  where the sublattice and valley degrees of freedom coincide. The flux attachment scheme, described in Sec. II, determines the order parameters. Our aim is  to understand how the collective excitations change in the presence of the order parameters. This is potentially a path to gain insight into the nature of symmetry breaking present in the ZLL in graphene. \par
Our main results are: i) We find expressions allowing calculation of the collective excitation spectra for FQH states in graphene in the presence of symmetry breaking orders; ii) we find that states with the simplest flux attachments are generally the most stable in that a) they have larger gaps as momentum $q \rightarrow 0$ and b) the magnetoroton minima at finite $q$ have larger energies;  iii) We investigated the excitation spectra for $\nu = 1/3$, $\nu = 1/2$ and $\nu = 2/5$ states while varying order parameters and flux attachments. We find that there is a complex interplay of these two factors that may have implications for recent experimental measurements \cite{Zibrov2018}. \par 
 This paper is structured as follows. In Sec.\ II we introduce the model. In Sec.\ III we derive the effective action and look for the saddle point configuration then expand the mean field action around the saddle point in terms of Gaussian fluctuations. In Sec. IV we derive an expression for the electromagnetic response tensor. In Sec. V we present numerical results for the collective 
 excitation spectra and in Sec. VI we discuss our results and conclude.

\section{Model} 
Starting from the extended Hubbard model on the honeycomb lattice, and applying the Hartree-Fock approximation {\cite{Herbut2007b} gives rise to a low energy Hamiltonian, in sublattice space, including CSB orders $m_{\alpha}$ and ferromagnetic order $f_{\alpha}$ of  
\begin{equation}~\label{eqn:1}
H^{\xi}_{\alpha} = \xi_{\alpha} \hbar v_{F} \left( \Pi_{1}\sigma_{1} + \Pi_{2}\sigma_{2}\right) + m_{\alpha}\sigma_3 + f_{\alpha}\sigma_0,
\end{equation}
where $\Pi_{i}= p_{i} + e A_{i}; \smallskip i=1,2$, with $p_{i}$ the momentum operator and $A_{i}$ the vector potential. The index $\alpha=1,2,3,4$ labels components of the spin and valley degrees of freedom (also called flavours or species) as $1 \equiv K$$\uparrow$, $2 \equiv K$$\downarrow$, $3 \equiv K'$$\uparrow$ and $ 4 \equiv K'$$\downarrow$. $K$ and $K'$ are the two inequivalent Brillouin-zone (BZ) points where the valence band touches the conduction band in reciprocal lattice space. The sigma matrices act in the $2 \times 2$ sublattice space and $\xi_{\alpha}=\pm$ correspond to the $+(K)$ and $-(K')$ valleys respectively. The Hamiltonian [Eq.~$\eqref{eqn:1}$] acts on the spinor $\Psi_{\alpha} = (u_{\alpha}, v_{\alpha})^{T}$ where $u_{\alpha} (v_{\alpha})$ is the amplitude on the $A (B)$ sublattice of graphene's honeycomb lattice. \par
In Eq.~$\eqref{eqn:1}$ the $m_{\alpha}$ are a combination of chiral symmetry breaking orders defined as: $m_{1} = C +N; m_{2} = C - N; m_{3} = - (C + N); m_{4} = - (C - N)$ where C is the charge density wave order and N is easy-axis Neel anti-ferromagnetic order. The ferromagnetic order (F) enters Eq.~$\eqref{eqn:1}$ through $f_{\alpha}$, defined as: $f_{1} = F; f_{2} = - F ; f_{3} = F ; f_{4} = - F $.

Equation~$\eqref{eqn:1}$ describes interacting electrons in graphene in the presence of a magnetic field at the mean field level. A system of electrons in a magnetic field can be equivalently described by a system of composite fermions (CFs) in an effective magnetic field \cite{Jain}. We consider four different species or flavours of composite fermions, corresponding to the different values of $\alpha$ as defined above. We begin by introducing the transformation $\Psi_{\alpha} = e^{i \Phi_{\alpha}}\psi_{\alpha}$, where $\psi_{\alpha}$ is the composite fermion field \cite{Narayanan2018}, and 
\begin{equation}
\Phi_{\alpha}  = \mathcal{K}_{\alpha \beta} \int d\vec{\rm{r'}} \rm{arg}(\vec{r} - \vec{r'})  \rho_{\beta}(\vec{r'}), 
\end{equation}
 where the matrix $\mathcal{K}$ describes the flux attachment scheme. An element $\mathcal{K}_{\alpha \beta}$ is the flux attached to CF of species $\alpha$ as seen by the species $\beta$. We parametrise $\mathcal{K}$ using the following form \cite{Modak2011}
 \begin{eqnarray}
\mathcal{K} = \left(\begin{array}{cccc} 
      2k_{1} & m_{1} &  n_{1} & n_{2} \\ 
      m_1 & 2k_{2} & n_3 & n_4 \\
      n_1 & n_3 & 2k_3 & m_2 \\
		  n_2 & n_4 & m_2 & 2k_4 
\end{array} \right).
\end{eqnarray} 
With the definition of the matrix element $\mathcal{K}_{\alpha \beta}$ as the flux attached to species $\alpha$ as seen by species $\beta$ the physical meaning of the elements $k$, $m$ and $n$ emerges: $2k$ is the flux attached to the each species as seen by itself, $m_i$ is the flux attached to a species as seen by another species that belongs to the same valley $K/K'$ but opposite spin and $n_i$ is the flux attached to a species, belonging to a valley $K/K'$ as seen by another species belonging to the other valley $K'/K$. 
For our calculations we consider the elements of $\mathcal{K}$ under the simplification $k_i=k$, $m_i=m$ and $n_i=n$ which are labelled by the triplet $(k,m,n)$. 

The derivative terms in the Hamiltonian transform as
\begin{equation}
\Psi^{\dagger}_{\alpha}(\pm \sigma_{1} \Pi_{1} - \sigma_{2} \Pi_{2}) \Psi_{\alpha} \rightarrow \psi_{\alpha}^{\dagger} (\pm \sigma_{1} \tilde{\Pi}_{1} - \sigma_{2} \tilde{\Pi}_{2})\psi_{\alpha}, \nonumber
\end{equation}
where $  \tilde{\Pi}_{i} = \Pi_{i} - a^{\alpha}_{i}$, with the Chern-Simons field $a^{\alpha}_{i}$ defined as
\begin{equation}
a^{\alpha}  = K_{\alpha \beta} \int  d\vec{r'} g(\vec{r} - \vec{r'})  \rho_{\beta}(\vec{r'}); \quad g(\vec{r}) = \frac{\hat{z} \times \vec{r}}{r^2}. \nonumber
\end{equation}
Here $\rho_{\alpha}$ corresponds to the density of composite fermion species of type $\alpha$. In terms of these densities we can define our order parameters as follows \cite{Frassdorf2018, Narayanan2018} 
\begin{eqnarray}~\label{eqn:2}
1 = \frac{\rho_1 + \rho_2 + \rho_3 + \rho_4}{\rho}, C = \frac{\rho_1 + \rho_2 - \rho_3 - \rho_4}{\rho}, \nonumber \\
F = \frac{\rho_1 - \rho_2 + \rho_3 - \rho_4}{\rho}, N = \frac{\rho_1 - \rho_2 - \rho_3 + \rho_4}{\rho}.
\end{eqnarray}

 The composite fermion Hamiltonian is thus
\begin{equation}~\label{eqn:3}
H_{\alpha}^{\xi} = \xi_{\alpha} v_F (\tilde{\Pi}^{\alpha}_{1}\sigma_{1} + \tilde{\Pi}^{\alpha}_{2}\sigma_{2}) + m_{\alpha}\sigma_3 + f_{\alpha}\sigma_0,
\end{equation}
where $\Pi_{i}^{\alpha} = p_i + e A_i + a^{\alpha}_{i}$, with $\alpha$ again labelling the species. 

Following Fr\"{a}{\ss}dorf \cite{Frassdorf2018}, we now employ the Schwinger-Keldysh technique \cite{Schwinger, Keldysh1965, Chou1985, Kamenevbook} to develop a field theoretic description of the multi-species composite fermions coupled to four  statistical $U(1)$ gauge fields, $a^{\alpha}$. In the Schwinger-Keldysh technique the time argument is promoted from a real variable to a complex variable corresponding to a contour-time and the correlation functions are defined as path-ordered products of the fields on the Schwinger-Keldysh contour. Since we work with an equilibrium system we take the reference time, $t_0$ on the contour to be in the infinite past, thereby reducing the kinetic equation solutions to well known equilibrium distributions. The  Schwinger-Keldysh technique leaves open the option to extend our theory to a finite temperature and non-equilibrium scenarios.
\begin{figure}[htb]~\label{fig:1}
 \includegraphics[width=8cm]{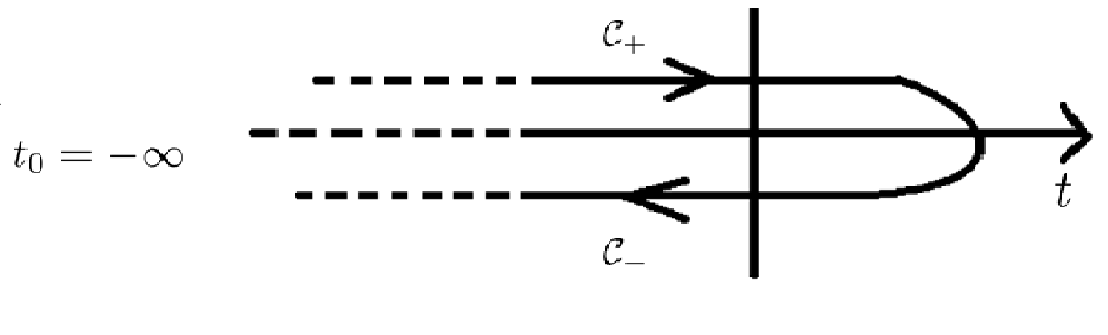}
 \caption{Schwinger-Keldysh closed contour with  forward time branch ($\mathcal{C}_+$) and backward time branch ($\mathcal{C}_-$). We set the reference time $t_0=-\infty$.}
\label{fig:Fig1}
\end{figure}

\section{Effective action}
The generating functional for the Hamiltonian defined in Eq.~$\eqref{eqn:1}$ is given by 
\begin{equation}
Z[\psi_{\alpha}, A_{\mu}, \mathcal{A}_{\mu}^{\alpha}, a_{\mu}^{\alpha}] = \int \mathcal{D}\psi^{\dagger}\mathcal{D}\psi e^{i S[\psi, e(A_{\mu} + \mathcal{A}_{\mu}^{\alpha}) + a_{\mu}^{\alpha}]} ,
\end{equation}
where the external vector potential, $A_{\mu} + \mathcal{A}_{\mu}^{\alpha}$, is composed of two terms: a  piece $A_{\mu}$ corresponding to the perpendicular magnetic field and a small fluctuating term with vanishing average, $\mathcal{A}_{\mu}^{\alpha} $, which is used to probe the electromagnetic response of the system. 

 The action $S$ can be written as
\begin{equation}~\label{eqn:5}
S = S_D + S_{CS},
\end{equation}
where $S_D$ is the composite fermion action corresponding to the Hamiltonian in Eq.~$\eqref{eqn:2}$: 
\begin{equation}
S_{D}[\psi, A_{\mu} + \mathcal{A}_{\mu}^{\alpha} + a_{\mu}^{\alpha}]  = \int_{C,\vec{r}} \psi^{\dagger,\alpha}\hat{G}_{0, \alpha \beta}^{-1} \psi^{\beta} ,
\end{equation}
with
\begin{equation}
\int_{C,\vec{r}} \equiv \int_{C} dt \int d^{2}r,
\end{equation}
and C is the Schwinger-Keldysh contour along which the integration is performed. The matrix $\hat{G}_{0}^{-1}$ is the inverse contour-time propagator which is diagonal in the species index:
\begin{equation}
\hat{G}^{-1}_{0}  = \mathrm{diag} (G^{-1}_{0,K\uparrow}, G^{-1}_{0,K\downarrow},G^{-1}_{0,K'\uparrow},G^{-1}_{0,K'\downarrow}),
\end{equation}
with
\begin{equation}
G^{-1}_{0,\alpha} (x,y) = \delta_C (x - y) (i \sigma^{\mu}_{\alpha} \mathcal{D}^{\alpha}_{\mu} - \mu_{\alpha} + m_{\alpha}\sigma_3 + f_{\alpha} \sigma_0 ).
\end{equation}
We note that we have already included interaction terms at the mean field level, which leads to order parameters C and N (that combine to form $m_{\alpha}$) and F. 
Here $\delta_C (x - y) = \delta_C (x_0 - y_0) \delta( \vec{x} - \vec{y})$ is the contour-time delta function and $\sigma^{\mu}_{\alpha} = (\sigma_0, \kappa_{\alpha}v_F \sigma_1, \kappa_{\alpha} v_F \sigma_2)$ and $m_{\alpha}, f_{\alpha}$ have been defined above. The gauge covariant derivative
\begin{equation}
\mathcal{D}^{\alpha}_{\mu} = \partial_{\mu} + i e A_{\mu} + i e \mathcal{A_{\mu}^{\alpha}} + i a_{\mu}^{\alpha}, 
\nonumber
\end{equation}
contains the fields $  A_{\mu} + \mathcal{A_{\mu}^{\alpha}}$ and the statistical gauge field $a_{\mu}^{\alpha}$.

$S_{CS}$ is the Chern-Simons action which describes the dynamics of the statistical gauge field $a_{\mu}^{\alpha}$ and has the  form
\begin{equation}
S_{CS} = \frac{1}{2} (\mathcal{K})^{-1}_{\alpha \beta} \int_{C,\vec{r}} \epsilon^{\mu \nu \lambda} a_{\mu}^{\alpha}\partial_{\nu} a_{\lambda}^{\beta}.
\end{equation}

We integrate out the fermionic fields $\psi$ from the action $S$ in Eq.~$\eqref{eqn:5}$  to obtain an effective action in terms of the gauge fields only, 

\begin{equation}~\label{eqn:9}
\begin{split}
S_{\rm{eff}}[ e(A_{\mu} + \mathcal{A}_{\mu}^{\alpha}) , a_{\mu}^{\alpha}] &= -i \mathrm{Tr\enskip ln} G_{0}^{-1} \left[ e(A_{\mu} + \mathcal{A}_{\mu}^{\alpha}) , a_{\mu}^{\alpha}\right] \\ & \quad \quad \quad+ S_{CS}\left[a_{\mu}^{\alpha}\right].
\end{split}
\end{equation}
We find the saddle point configuration of the path integral for the statistical gauge fields $a^{\alpha
}_{\mu}$ and then perform an expansion of the effective action in terms of fluctuations around this mean field solution. Following Fradkin and Lopez \cite{Lopez1991} we search for a solution that leads to a vanishing charge carrier current and a non-zero, time independent charge carrier density, $\rho_{\alpha}$, which is given by 
\begin{equation}
\rho_{\alpha} = - (\mathcal{K})^{-1}_{\alpha \beta}\mathcal{B}^{\beta} ,
\end{equation}
where $\mathcal{B}^{\beta}$ is a uniform field due to the statistical gauge field experienced by  a CF of species $\beta$. Inverting this relation gives us
\begin{equation}
\mathcal{B}^{\beta} = - \rho_{\alpha} (\mathcal{K})^{\alpha \beta}.
\end{equation}

The effect of this field is to reduce/enhance (depending on the sign of charge carriers present in the sample) the original magnetic field so that a CF of species $\alpha$ experiences an effective magnetic field given by
\begin{equation}~\label{eqn:B}
B_{\rm{eff}}^{\alpha} = B + \mathcal{B}^{\alpha} = B - \rho_{\beta}(\mathcal{K})^{\alpha \beta}.
\end{equation}
Here $\rho$ is the total electron density and $\nu$ is the filling fraction for the electrons. From Eq. $\eqref{eqn:B}$  we get a relationship  \cite{Modak2011}  between the composite fermion filling fraction $\nu_{\alpha}$ for species $\alpha$, the density $\rho_{\alpha}$ and the flux attachment matrix of
\begin{equation}
\frac{\rho_{\alpha}}{\nu_{\alpha}} = \frac{\rho}{\nu} - \mathcal{K}^{\alpha \beta} \rho_{\beta}.
\end{equation}

 We now represent the effective action [Eq.~$\eqref{eqn:9}$] in a more convenient form by performing a Keldysh rotation. The contour illustrated in Fig.~\ref{fig:Fig1} consists of a forward ($\mathcal{C}_+$) and a backward ($\mathcal{C}_-$) piece, and the fields on the respective pieces of the contour maybe be written as $\psi_{\pm}$, $a_{\pm}$. We transform to a new set of double fields $\psi_{c,q}$ and $a_{c,q}$ which are symmetric and antisymmetric linear combinations of the $\pm$ double fields with, e.g. for $\psi$:

 \begin{equation}
 \psi_{c} = \frac{1}{\sqrt{2}}\left(\psi_+ + \psi_- \right), \quad \psi_q =  \frac{1}{\sqrt{2}}\left(\psi_+ + \psi_- \right).
 \end{equation}
 The labels $c,q$ correspond to classical and quantum components respectively \cite{Lozano}.  The net result is that the derivatives of the action with respect to the gauge fields
are now performed with respect to the quantum components \cite{Chou1985}.
\par
The gauge fields, $a^{\alpha}$, can be viewed as being comprised of a mean field part ($\bar{a}^{\alpha}$) and a fluctuation part ($\Delta a^{\alpha}$), $a^{\alpha} = \bar{a}^{\alpha} + \Delta a^{\alpha}$, and we expand the effective action in terms of the fluctuations up to the second order in $\Delta a$. Terms linear in fluctuations vanish and we get 
 \begin{widetext}

\begin{flalign}
S_{\rm{eff}}[ \mathcal{A}_{\mu}^{\alpha}, a_{\mu}^{\alpha}] = \int_{xy} &\left[(\Delta a_c)^{\alpha}_{\mu} + ( \mathcal{A}_c)_{\mu}^{\alpha}) \enskip (\Delta a_q)^{\alpha}_{\mu} + ( \mathcal{A}_q)_{\mu}^{\alpha}\right] \left(x\right)  \left[\begin{array}{cc} 
    0 & (\Pi^A)^{\mu \nu}_{\alpha\beta} \\
   (\Pi^R)^{\mu \nu}_{\alpha\beta} & (\Pi^K)^{\mu \nu}_{\alpha\beta} 
\end{array} \right]\left(x,y\right) \left[\begin{array}{c} (\Delta a_c)^{\beta}_{\nu}
+ ( \mathcal{A}_c)_{\nu}^{\beta}) \\
 (\Delta a_q)^{\beta}_{\nu} + ( \mathcal{A}_q)_{\nu}^{\beta})
\end{array} \right]\left(y\right) \\  + & \left[(\Delta a_c)^{\alpha}_{\mu} \enskip (\Delta a_q)^{\alpha}_{\mu} )\right]\left(x\right) \left[\begin{array}{cc} 
    0 & (C^A)^{\mu \nu}_{\alpha\beta} \\
   (C^R)^{\mu \nu}_{\alpha\beta} & (C^K)^{\mu \nu}_{\alpha\beta} 
\end{array} \right]\left(x,y\right) \left[\begin{array}{c} (\Delta a_c)^{\beta}_{\nu} \\ (\Delta a_q)^{\beta}_{\nu} \end{array} \right]\left(y\right) , \nonumber
\end{flalign}
\end{widetext} 
which can be written in a more compact form as :
 \begin{widetext}
  \begin{equation}
 S_{\rm{eff}}[\mathbf{\mathcal{A}}_{\mu}^{\alpha},\mathbf{a}_{\mu}^{\alpha}] = \int_{xy} \left\{ \left[ \mathbf{(\Delta a)}^{\alpha}_{\mu} +  \mathbf{(\mathcal{A}})_{\mu}^{\alpha})\right]^{\mathbf{T}}(x) \mathbf{\Pi}_{\alpha \beta}^{\mu \nu}(x,y) \left[ (\bf{\Delta a)^{\beta}_{\nu}} + ( \mathbf{\mathcal{A})_{\nu}^{\beta}}\right](y) +  (\mathbf{\Delta a})^{\alpha}_{\mu}(x) \mathbf{C_{\alpha \beta}^{\mu \nu}} (x,y)  (\mathbf{\Delta a})^{\beta}_{\nu} (y)\right\}.
   \end{equation}
\end{widetext} 
Here the fields $\bf {\mathcal{A}_{\mu}^{\alpha}}$ and $\bf{ a_{\mu}^{\alpha}} $ are two component fields in Keldysh space
\begin{equation}
{\mathbf{\mathcal{A}}} = \left(\begin{array}{c} \mathcal{A}^{c} \\ \mathcal{A}^{q} \end{array} \right), \nonumber
\end{equation}
and similarly for the fields $\bf{a^{\mu}_{\alpha}}$.
 The polarization tensor ${\bf \Pi}$  and the Chern-Simons tensor ${\bf C}$ are $2 \times 2$ matrices with advanced (A), retarded (R) and Keldysh (K) components:
 \begin{equation} 
\bf{C}^{\mu \nu}_{\alpha \beta} = \left(\begin{array}{cc} 0 & (C^{A})^{\mu \nu}_{\alpha \beta}\\
 (C^{R})^{\mu \nu}_{\alpha \beta} &  (C^{K})^{\mu \nu}_{\alpha \beta} \end{array}\right).
 \end{equation}
 
We use bold font to indicate that a quantity has classical (c) and quantum (q) components if a vector, or Advanced (A), Retarded (R) and Keldysh (K) components if a $2 \times 2$ matrix.
 
As the system is in equilibrium, in the linear response regime, the different components satisfy the bosonic fluctuation-dissipation theorem. The polarization tensor $\bf{\Pi}$ is given by
\begin{equation}
{\bf \Pi}_{\alpha\beta}^{\mu \nu} = - \frac{i}{2} \frac {\delta^2}{\bf{\delta a^{\beta}_{\nu} \delta a^{\alpha}_{\mu}}}\left.\mathrm{Tr} \enskip {\rm ln} \enskip {\bf \hat{G}_{0}^{-1}} [e{\bf A_{\mu}} +\bf{a_{\mu}^{\alpha}}]\right|_{\bf{a = \bar{a}}},
\end{equation}
where ${\bf A_{\mu}}$ is the electromagnetic field and  $\bf{\hat{G}_{0}^{-1}}$ is the inverse time propagator mapped onto the Keldysh basis. Since the propagators are diagonal in the species index $\alpha$, the polarization tensor is also diagonal,  ${\bf{\Pi^{\mu \nu}_{\alpha \beta}}} = {\bf{\Pi_{\alpha \beta}^{\mu \nu} \delta^{\alpha \beta}}}$. To determine the collective excitation spectra we only need consider the retarded (R) components of the polarization tensor given in Appendix A. 

The Chern-Simons tensor has the following form 

\begin{equation}
C^{\mu \nu}_{\alpha \beta} = (\mathcal{K})^{-1}_{\alpha \beta} \epsilon^{\mu \nu \lambda}\partial_{\lambda}.
\end{equation}

The polarization tensor and the Chern-Simons tensor are transverse. As a consequence of this the polarization tensor can be decomposed into scalars $\bf{\Pi}^0$, $\bf{\Pi}^{1}$ and $\bf{\Pi}^{2}$ \cite{Lopez1991, Lopez1993} as follows:
\begin{eqnarray}{\label{eqn:{17-22}}}
{\bf{\Pi}}_{\alpha \beta}^{00}(\omega, {\bf{q}}) &=& -q^2 {\bf{\Pi}}_{\alpha \beta}^{0}, \\
{\bf{\Pi}}_{\alpha \beta}^{0i}(\omega, {\bf{q}}) &=& -\omega q^{i}{\bf{ \Pi}}_{\alpha \beta}^{0}(\omega, {\bf{q}}) + i \epsilon^{0ij}q_{j} {\bf{\Pi}}_{\alpha \beta}^{1}(\omega, {\bf{q}}),\\
{\bf{\Pi}}_{\alpha \beta}^{i0}(\omega, {\bf{q}}) &=& -\omega q^{i} {\bf{\Pi}}_{\alpha \beta}^{0}(\omega, {\bf{q}}) - i \epsilon^{0ij}q_{j}{\bf{ \Pi}}_{\alpha \beta}^{1}(\omega, {\bf{q}}),\\ \nonumber
{\bf{\Pi}}_{\alpha \beta}^{ij}(\omega, {\bf{q}}) &=& - \omega^2 \delta^{ij}  {\bf{\Pi}}_{\alpha \beta}^{0}(\omega, {\bf{q}})  + i \epsilon^{0ij}\omega {\bf{\Pi}}_{\alpha \beta}^{1}(\omega, {\bf{q}})\\  &  & +( \delta^{ij} {q}^2 - q^{i}q^{j})  {\bf{\Pi}}_{\alpha \beta}^{2}(\omega, {\bf{q}}).
\end{eqnarray}
\section{Electromagnetic response tensor} 
In order to obtain the electromagnetic response tensor we integrate over the statistical gauge fields. Due to the transverse nature of the polarization and Chern-Simons tensor the inverse of both is ill defined and so is the inverse of the sum of these two tensors, $(\bf{\Pi + C})^{-1}$ , which appears when we perform the integration over the statistical gauge fields. 

In order to overcome this problem one can employ the Fadeev-Poppov method \cite{Peskin}. The result of this is a gauge fixed generating functional of the form
\begin{equation}
\mathcal{Z}_{\rm{GF}}[\mathcal{A_{\mu}^{\alpha}}]  = \int \left(\mathcal{D}\Delta a \right) \enskip e^{i (S_{\rm{eff}}[\mathcal{A}, \Delta a] +S_{\rm{GF}}[\Delta a])},
\end{equation}
where the gauge fixing action has the form
\begin{equation}
S_{\rm{GF}} =\left(\frac{1}{2 \eta}\right) \int_{\mathcal{C},x} (\partial_{\mu} \Delta a^{\mu})^2= \frac{1}{2} \int_{\mathcal{C},x} \Delta a^{\mu} \mathcal{G}_{\mu \nu} \Delta a^{\nu},
\end{equation} 
where $\eta$ is a real valued parameter which we can set to be $\eta =1$. Since the electromagnetic tensor is a physical quantity the choice of gauge should not matter and hence all references to the parameter $\eta$ drop out in the end. Now we can perform the integral over the gauge fields since the addition of $\mathcal{G}$ makes the sum $\bf{\Pi + \mathcal{G} +C}$ invertible. The object that we obtain as a consequence of performing the integral is the electromagnetic response tensor which has the form
\begin{equation}~\label{eqn:28}
\bf{K = \Pi - \Pi (\Pi + \mathcal{G} + C)^{-1} \Pi}.
\end{equation}
   The electromagnetic tensor can be expressed, similarly to the polarization and Chern-Simons tensors, in Keldysh space as a $2 \times 2$ matrix with advanced (A), retarded (R) and Keldysh (K) components: 
\begin{equation} 
\bf{K} = \left(\begin{array}{cc} 0 & (K^{A})^{\mu \nu}_{\alpha \beta}\\
 (K^{R})^{\mu \nu}_{\alpha \beta} &  (K^{K})^{\mu \nu}_{\alpha \beta} \end{array}\right).
 \end{equation}
The Keldysh component is related to the advanced and retarded components through the bosonic fluctuation-dissipation theorem:
 \begin{equation}
 K^{K}_{\omega ,\bf{q}} = \coth{\left(\frac{\omega}{2T}\right)}( K^{R}_{\omega ,\bf{q}} -  K^{A}_{\omega ,\bf{q}}).
 \end{equation}

 The electromagnetic response tensor is also transverse and hence admits a decomposition, similar to the polarization tensor, in terms of scalar kernels, $K_0$, $K_1$ and $K_2$ which can be written as 
 \begin{eqnarray}~\label{eqn:33}
 K_{0}^{R}(\omega, {\bf{q}}) &=& -(\mathcal{K}^{-1})^2 \frac{\Pi^{R}_{0}(\omega, {\bf{q}})}{D^{R}(\omega, {\bf{q}})}, \\ 
  K_{1}^{R}(\omega, {\bf{q}}) &=& \mathcal{K}^{-1} + (\mathcal{K}^{-1})^2 \frac{(\mathcal{K}^{-1}+\Pi^{R}_{1}(\omega,  {\bf{q}}))}{D^{R}(\omega,  {\bf{q}})}, \\
 K_{2}^{R}(\omega,  {\bf{q}}) &=&  (\mathcal{K}^{-1})^2  \frac{\Pi^{R}_{2}(\omega, {\bf{q}})}{D^{R}(\omega, {\bf{q}})}.
 \end{eqnarray}
 Since the polarization tensor is diagonal in the species index $\alpha$, it commutes with $\mathcal{K}^{-1}$. Here $D^{R/A}$ is the denominator matrix which has the form
 \begin{equation}
 \label{eqn:36}
D^{R/A} = \omega^2 (\Pi^{R/A}_{0})^2 -  (\mathcal{K}^{-1}+\Pi_1^{R/A})^2 + q^2 (\Pi_0^{R/A} \Pi_2^{R/A}).
 \end{equation}
 The retarded and advanced kernels are Hermitian conjugates of each other. Full expressions for $\Pi_0$, $\Pi_1$, $\Pi_2$ are given in Appendix A.  \par
 The denominator matrix $D$ is of central importance to our work. The zeros of the determinant of the denominator matrix gives us the location of poles for the electromagnetic response tensor.

 \begin{figure*}[htb]
\includegraphics[width=16cm]{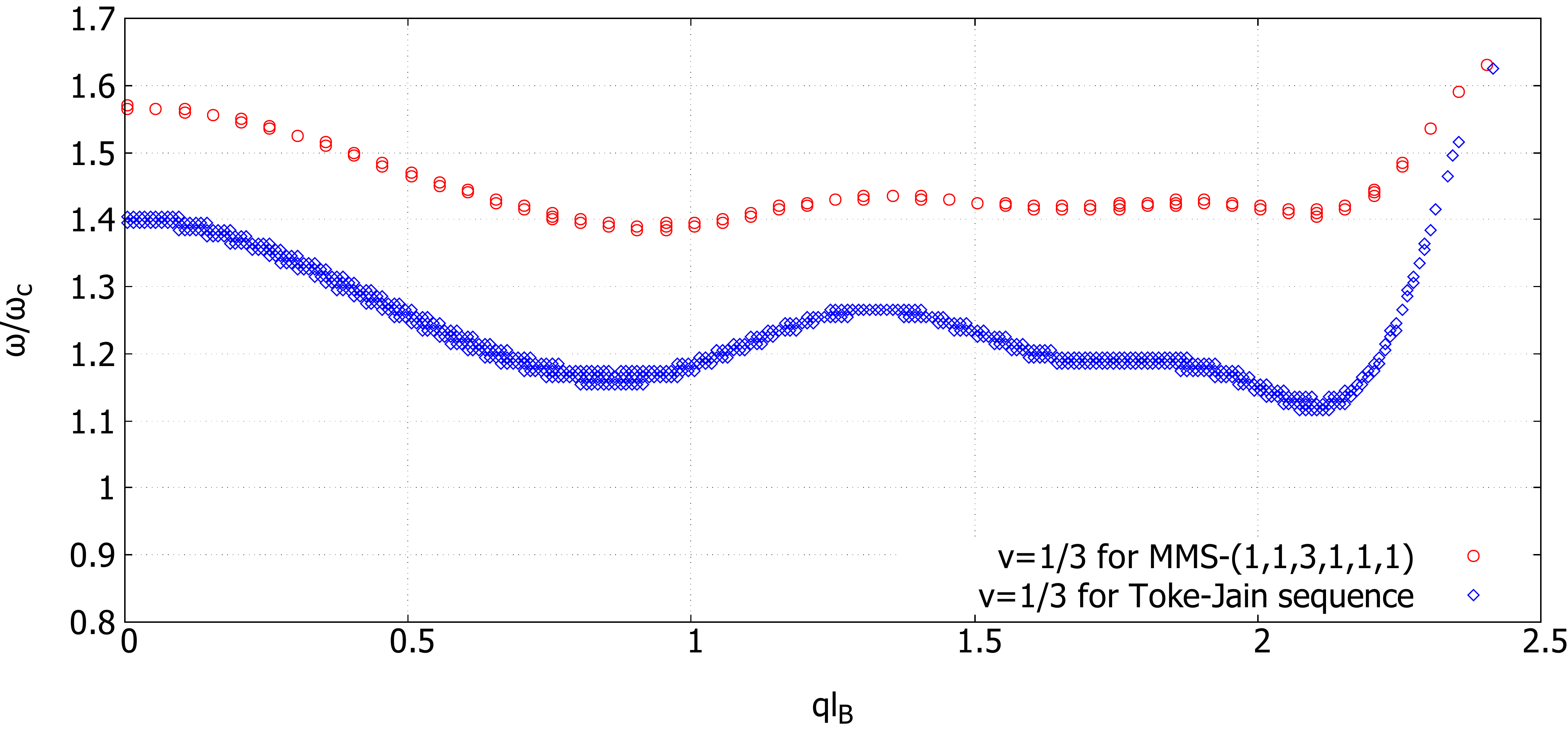}\quad
\caption{Location of poles as a function of $\frac{\omega}{\omega_c}$ and $ql_B$ for $\nu =1/3$ for the Toke-Jain sequence (black) and the MMS state for $(k,m,n) = (1,1,3)$ (red). Here $\omega_c = \sqrt{2}\frac{v_F}{l_B}$}
\label{fig:TJvsMMS}
\end{figure*}

\begin{figure*}[htb]
\includegraphics[width=16cm]{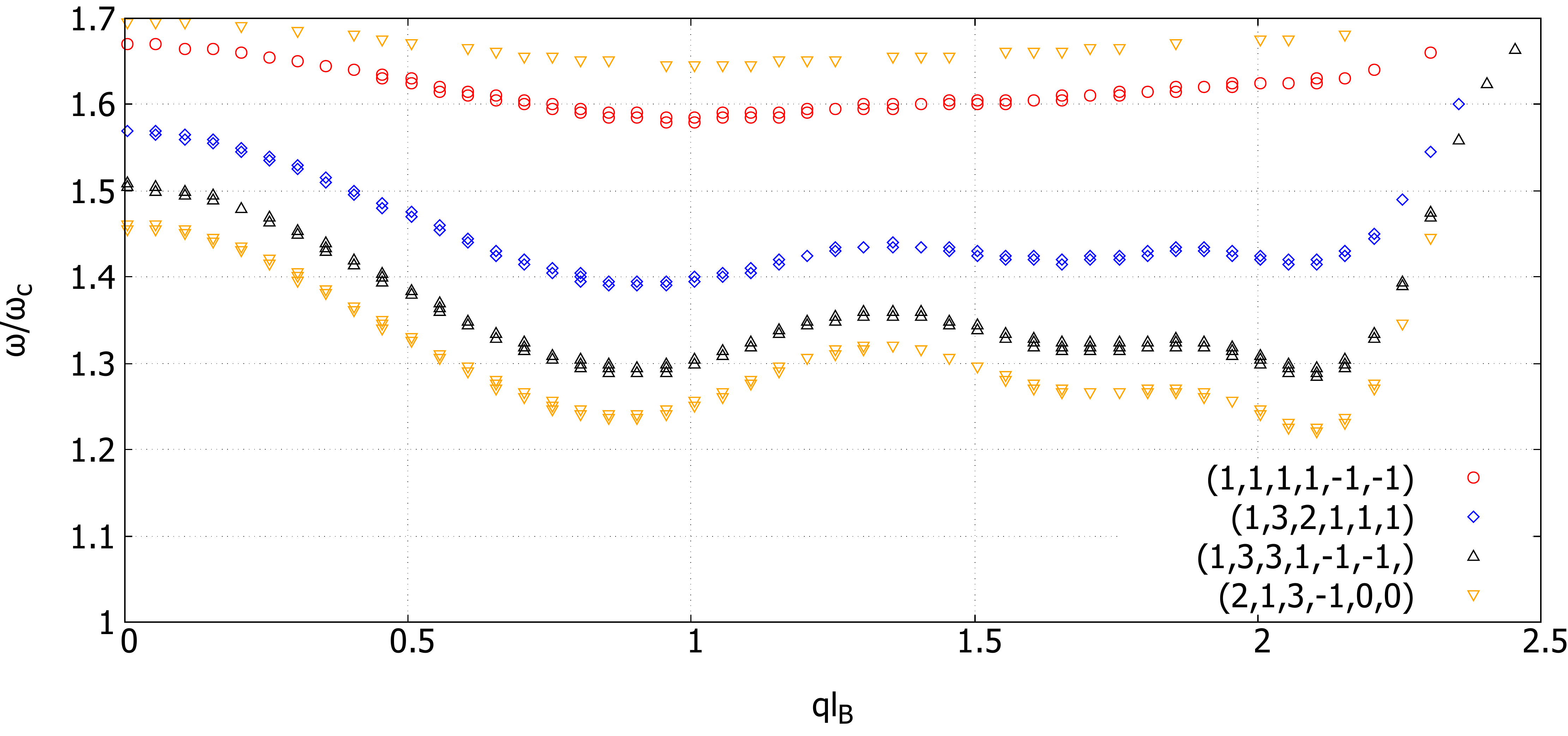}\quad
\caption{Location of poles as a function of $\frac{\omega}{\omega_c}$ and $ql_B$ for $\nu =1/3$ for the parameter set defined in the text $(k,m,n,C,N,F)$. $(1,1,1,1,-1,-1)$ (red), $(1,3,2,1,1,1)$ (blue), $(1,3,3,1,-1,-1)$ (black) and $(2,1,3,-1,0,0)$ (orange).}
\label{fig:MMSstates}
\end{figure*}

\section{Numerical Results}
In this section we find the poles of the electromagnetic response tensor numerically so as to determine the collective excitation spectra of various FQH states in graphene. Specifically, we calculate the zeros of the determinant of the denominator matrix $D^{R/A}$ given in Eq. $\eqref{eqn:36}$. We calculate the excitation spectra for flux attachment schemes at several different filling fractions $\nu$.  \par
Different flux attachment schemes are encoded in the matrix elements $\mathcal{K}_{\alpha \beta}$. 
Using the parametrization presented in Eq.~$\eqref{eqn:3}$, the simplest case is when the same number of flux quanta, $2k$, 
is attached to all the species \cite{Csaba2007, Modak2011}. 
This is the same as considering a single species with $2k$ flux quanta attached to it. 
For this case the filling fractions of FQH states are given by $\nu = \frac{\nu^*}{2k\nu^*+1}$ 
with $\nu^* = \nu_1 + \nu_2 + \nu_3 + \nu_4$, which is known as the T{\"{o}}ke-Jain sequence \cite{Csaba2007}. 
Here $\nu_{\alpha}$, $\alpha = 1,2,3,4$, are the filling fractions of the individual species. 
Hence the T{\"{o}}ke-Jain sequence can be characterized by the set of parameters $(\nu^*, k)$. Following Refs. \cite{Modak2011, Narayanan2018},
we also consider the following simplification of $\mathcal{K}$: $k_i=k$, $m_i=m$ and $n_i=n$ for all $i$, which we refer to as the Modak-Mandal-Sengupta (MMS) states. This allows us to label MMS flux attachment schemes by the triplet $(k,m,n)$. In order to specify a FQH state with a given filling $\nu$, we also need to specify fillings $\nu_{\alpha}$ for the composite fermion Landau levels. Once $\nu_{\alpha}$, $\nu$ and $(k,m,n)$ are specified this determines the values of the order parameters $C$, $N$ and $F$ \cite{Modak2011, Narayanan2018}. We characterize the states we consider by the flux attachment parameters $(k,m,n)$ determined from the $\mathcal{K}$ matrix. In our numerical calculations we truncated the number of Landau levels (labelled by n and n$^\prime$) included in the calculation of $D$ to $N_c=10$ in Eq. $\eqref{eqn:A2}$ in order to cut computational time. We confirmed that our results were not sensitive to this choice of cut-off.

For each parameter set we calculate the collective modes and plot their angular frequency $\frac{\omega}{\omega_{c}}$ against $q l_B$ where $\omega_c =\sqrt{2} v_F/l_B $ is the cyclotron frequency with $v_F$ being the Fermi velocity and $l_B$ being the magnetic length. We characterize each dispersion curve by the following parameters, $\omega_g$: the energy gap as $q \rightarrow 0$; $q^i_{rm}$: the position of the $i^{th}$ magnetoroton minimum; and $\Delta^i_r$: the energy at the position of the $i^{th}$ magnetoroton minimum. We observe the following generic features in the dispersion curves for the lowest energy collective mode: i) a gap as $q \rightarrow 0$ and ii) minima for $ql_B \sim 1$, and $ql_B \sim 2.1$, which we attribute as a magneto-roton minima, similar to those seen for non-relativistic FQH states. Numerous states also have an additional minimum or shoulder for $ql_B \sim 1.7$.  
For higher energy collective modes, the dispersion is relatively flat in comparison to the lowest energy mode. 

In Fig.~$\ref{fig:TJvsMMS}$ we compare the dispersion curves for two different $\nu=1/3$ states, the T{\"{o}}ke-Jain state for the parameter set $(\nu^*=1, k=1)$ and the MMS state for the parameter set $(k=1, m=1, n=3)$. We notice that the MMS state has a higher gap as $q \rightarrow 0$ and has a higher magnetoroton energy as compared to the T{\"{o}}ke-Jain state. In the rest of the results we focus on MMS states motivated by their relevance for EDFQH states \cite{Narayanan2018}.

We consider several different MMS $\nu=1/3$ states and in Fig.~$\ref{fig:MMSstates}$ observe that: 
i) the roton energy is lowest for the state with order parameters $(C,F,N) = (1,-1,-1)$ and $(k, m, n) = (1, 1, 1)$; and
ii) as we increase $k$, $m$ or $n$ the energy of the roton, $\Delta_r$, decreases. 
In addition, we also observe the appearance of a second minimum around $ql_B \sim 2.1$ and what is sometimes a shoulder and sometimes a local minimum at around $ql_B \sim 1.7$.
Numerical values are tabulated in Table~\ref{tab:table1}.
  The minima deepen with increasing values of $k, m,$ and $n$.

\begin{widetext}

\begin{table}[H]
\caption{Dispersion parameters for selected $\nu = 1/3$ states}
\label{tab:table1}
\begin{center}
\begin{tabular}{|c|c|c|c|c|c|c|c|c|c|c|c|c|}
\hline
k & m & n & C & N & F & $\omega_g/\omega_c$ & $q_{rm}^1 l_B$ & $q_{rm}^2 l_B$ & $q_{rm}^3 l_B$ & $\Delta_r^1/\omega_c$ & $\Delta_r^2/\omega_c$ & $\Delta_r^3/\omega_c$ \\ \hline 
1 & 1 & 1 & 1 & -1 & -1 & 1.67 & 0.99 & - & - & 1.58 & - & -  \\
1 & 3 & 2 & 1 & 1 & 1 & 1.57 & 0.91 & 1.65 & 2.09 & 1.39 & 1.42 & 1.42 \\
1 & 3 & 3 & 1 & -1 & -1 & 1.51 & 0.90 & 1.73 & 2.10 & 1.29 & 1.32 & 1.29 \\
2 & 1 & 3 & -1 & 0 & 0 & 1.46 & 0.89 & 1.74 & 2.11 & 1.24 & 1.26 & 1.22\\
\hline
\end{tabular}
\end{center}
\end{table}

In Fig.~\ref{fig:varym} we show the variation of the position of the poles and the roton energy as we change $m$ in the triplet $(k,m,n)$. As before, the states we studied are parameterized by the set of parameters $(k,m,n,C,F,N)$. We fixed all the parameters except $m$. The results are summarized in Table \ref{tab:table2}.

\begin{table}[h]
\caption{Parameters for $\nu = 1/3$ states varying $m$}
\label{tab:table2}
\begin{tabular}{|c|c|c|c|c|c|c|c|c|c|c|c|c|}
\hline
k & m & n & C & N & F & $\omega_g/\omega_c$ & $q_{rm}^1 l_B$ & $q_{rm}^2 l_B$ & $q_{rm}^3 l_B$ & $\Delta_r^1/\omega_c$ & $\Delta_r^2/\omega_c$ & $\Delta_r^3/\omega_c$ \\ \hline 
1 & 0 & 2 & 1 & 1 & 1 & 1.65 & 0.98 & 1.66 & 2.08 & 1.53 & 1.56 & 1.58  \\
1 & 1 & 2 & 1 & 1 & 1 & 1.62 & 0.97 & 1.60 & 2.09 & 1.49 & 1.52 & 1.53
\\
1 & 3 & 2 & 1 & 1 & 1 & 1.57 & 0.91 & 1.65 & 2.09 & 1.39 & 1.42 & 1.42 \\
 
1 & 4 & 2 & 1 & 1 & 1 & 1.54 & 0.89 & 1.69 & 2.10 & 1.34 & 1.37 & 1.35 \\
\hline
\end{tabular}
\end{table}

In Fig.~\ref{fig:varyn}. we show the dispersions for several different $\nu=1/3$ MMS states with $k$ and $m$ fixed, but varying $n$. The results are summarized in Table \ref{tab:table3}.

\begin{table}[h]
\caption{Parameters for $\nu = 1/3$ states varying $n$}
\label{tab:table3}
\begin{tabular}{|c|c|c|c|c|c|c|c|c|c|c|c|c|}
\hline
k & m & n & C & N & F & $\omega_g/\omega_c$ & $q_{rm}^1 l_B$ & $q_{rm}^2 l_B$ & $q_{rm}^3 l_B$ & $\Delta_r^1/\omega_c$ & $\Delta_r^2/\omega_c$ & $\Delta_r^3/\omega_c$ \\ \hline 
1 & 1 & 1 & 1 & 1 & 1  & 1.67 & 0.99 & - & - & 1.58 & - & -  \\
1 & 1 & 2 & 1 & 1 & 1 & 1.62 & 0.97 & 1.60 & 2.09 & 1.49 & 1.52 & 1.53
\\
1 & 1 & 3 & 1 & 1 & 1 & 1.57 & 0.93 & 1.69 & 2.10 & 1.38 & 1.42 & 1.40 \\
 
1 & 1 & 4 & 1 & 1 & 1 & 1.50 & 0.90 & 1.68 & 2.13 & 1.29 & 1.31 & 1.25 \\
\hline
\end{tabular}
\end{table}

  \begin{figure}[H]
\includegraphics[width=16cm]{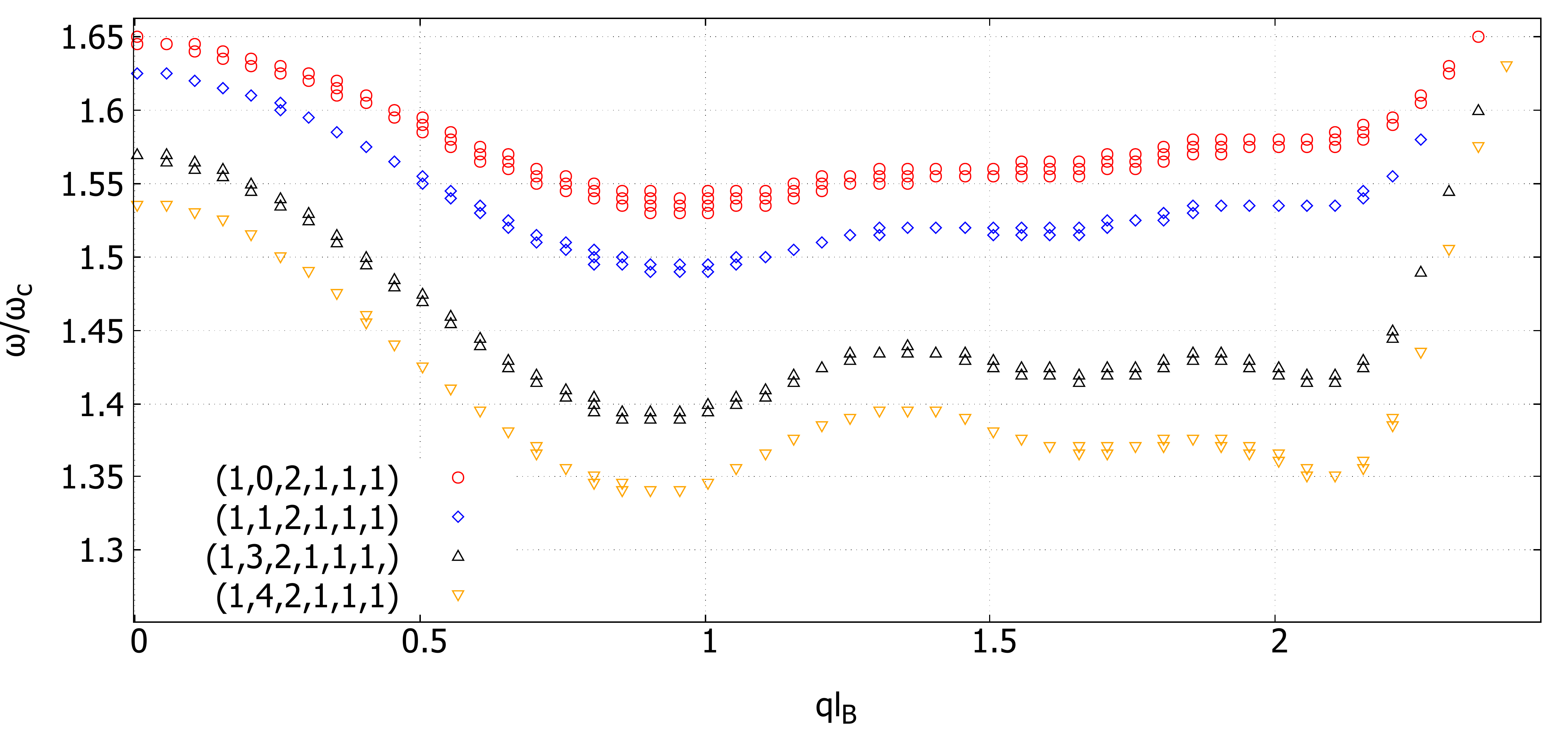}\quad
\caption{Dispersion curves for $\nu =1/3$ varying the parameter $m$. The parameter set defined in the text $(k,m,n,C,N,F)$. $(1,0,2,1,1,1)$ (red), $(1,1,2,1,1,1)$ (blue), $(1,3,2,1,1,1)$ (black), $(1,4,2,1,1,1)$ (orange).   }
\label{fig:varym}
\end{figure}

\end{widetext}
 
\begin{widetext} 
 
 \begin{figure}[H]
\includegraphics[width=16cm]{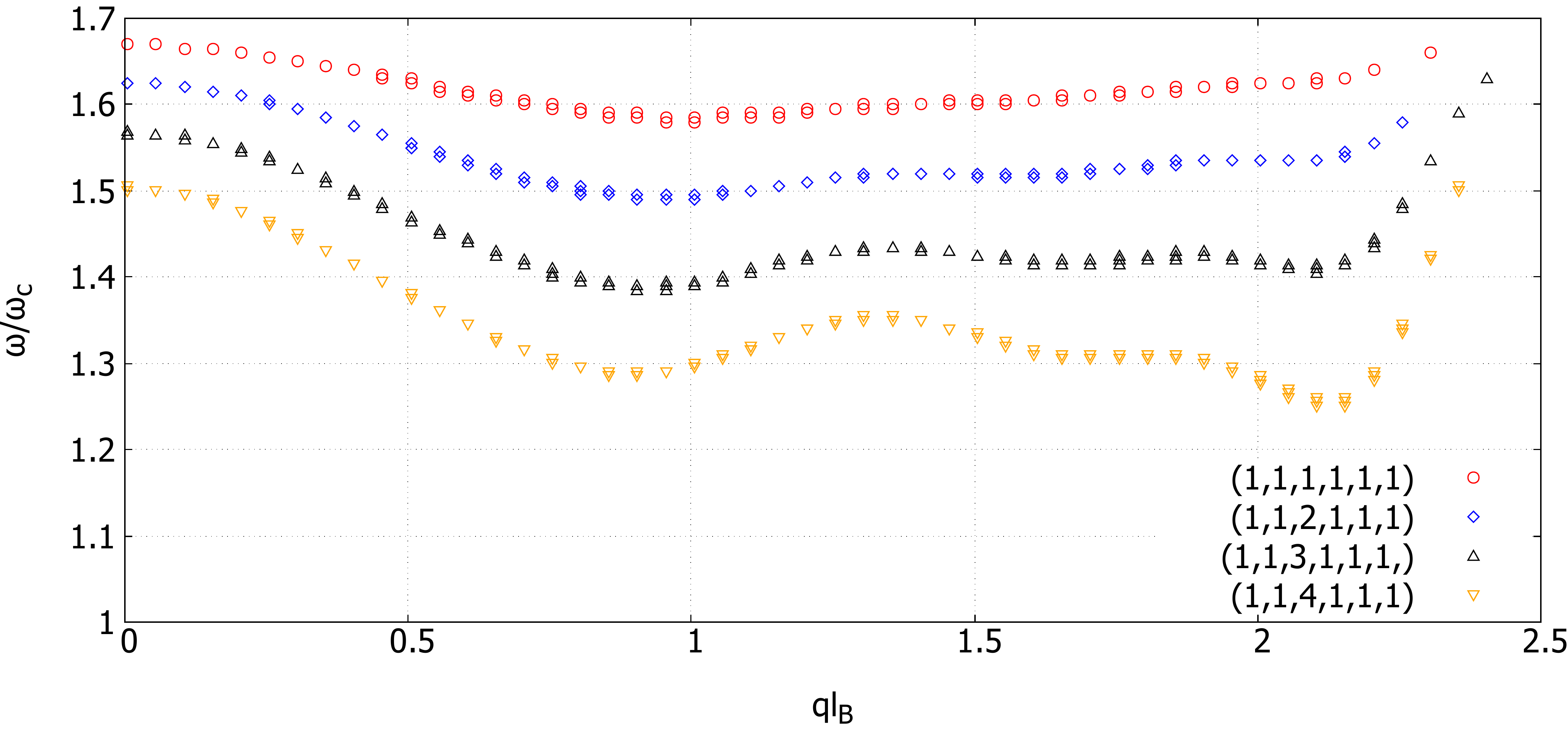}\quad
\caption{Dispersion curves for $\nu =1/3$ varying the parameter $n$. The parameter set defined in the text is $(k,m,n,C,N,F)$. $(1,1,1,1,1,1)$ (black), $(1,1,2,1,1,1)$ (red), $(1,1,3,1,1,1)$ (blue), $(1,1,4,1,1,1)$ (orange).}
\label{fig:varyn}
\end{figure}

Figures \ref{fig:varym} and \ref{fig:varyn}  lead to the following observations: i) the gap at $q \rightarrow 0$ decreases as we go from a low $m (n)$ value to a 
higher $m (n)$ value; ii) the position of the first roton minimum shifts towards slightly lower $q$ 
as we go from lower $m (n)$ to higher values of $m (n)$; iii) the position of the second and third roton minima shifts to slightly 
higher $q$ as we go from lower $m (n)$ to higher values; iv) the energies of the rotons $\Delta^i_r$ decreases as we go from lower $m (n)$ to higher values for $i = 1, 2, 3$.
In Tables \ref{tab:table2}  and \ref{tab:table3} we confirm that the observations i)-iv) hold as we go from lower $m (n)$ values to higher $m (n)$ values.

  \begin{figure*}[htb]
\includegraphics[width=16cm]{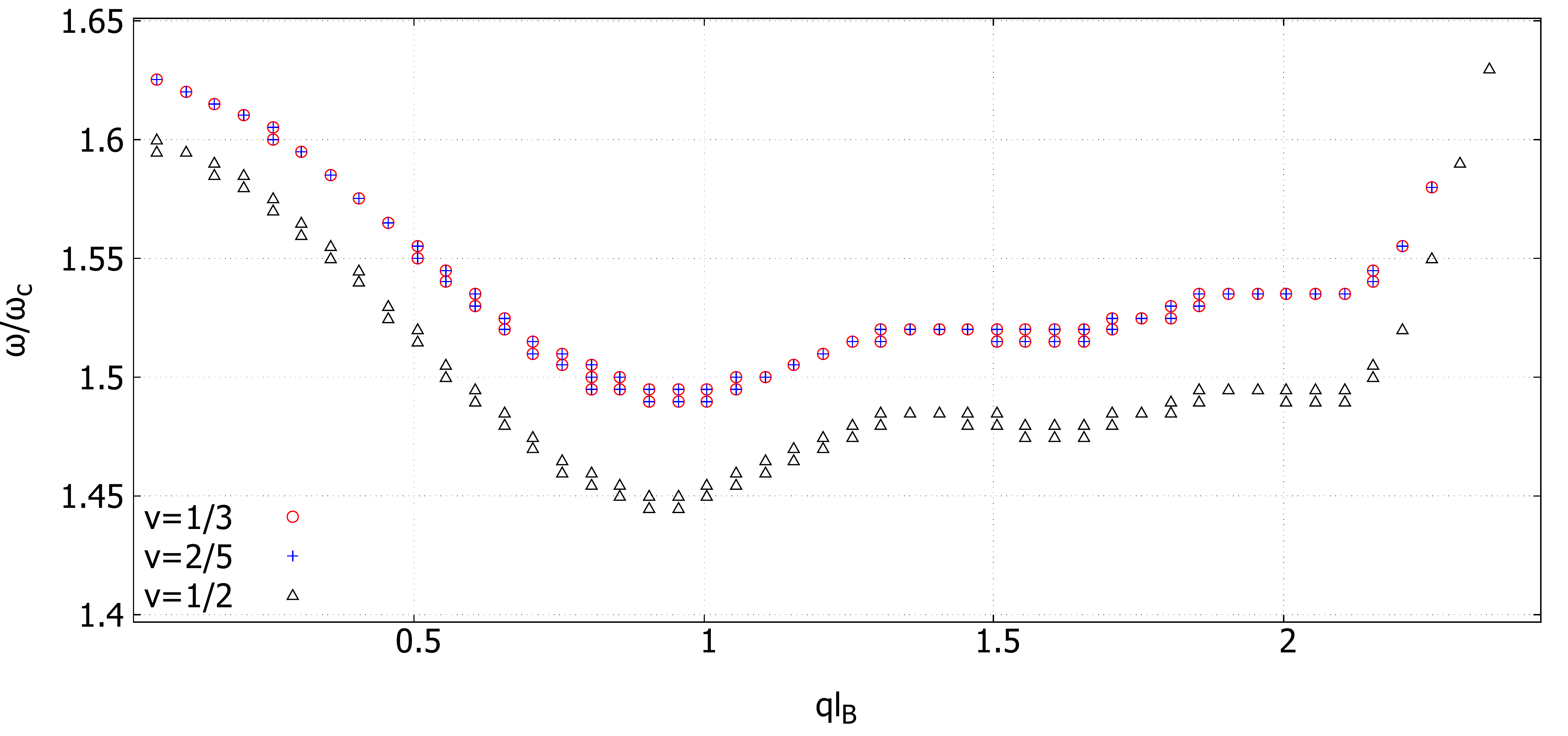}\quad
\caption{Dispersion curves for $\nu =1/3$ (1,1,2,1,1,1), $\nu=2/5$ (1,1,2,1,-1,-1) and $\nu=1/2$ (1,1,2,1,0,0). For this plot $(k,m,n)$ is the same for all three fractions while $(C,F,N)$ varies.    }
\label{fig:fixedkmn}
\end{figure*}

In addition to comparing the effect of different values of $m$ and $n$ on the collective mode spectrum for a given fraction, we also make a comparison of the collective mode spectra for several different incompressible fractions: $\nu =1/3$, $\nu =2/5$ and $\nu =1/2$. Unlike the single component case it is possible to have an incompressible FQH state for $\nu=1/2$ in a multicomponent quantum Hall system such as graphene \cite{Modak2011, Narayanan2018, Zibrov2018}. The relevant dispersion curves are plotted in Figs.~\ref{fig:fixedkmn} 
and \ref{fig:fixedCNF}.

\begin{figure*}[htb]
\includegraphics[width=16cm]{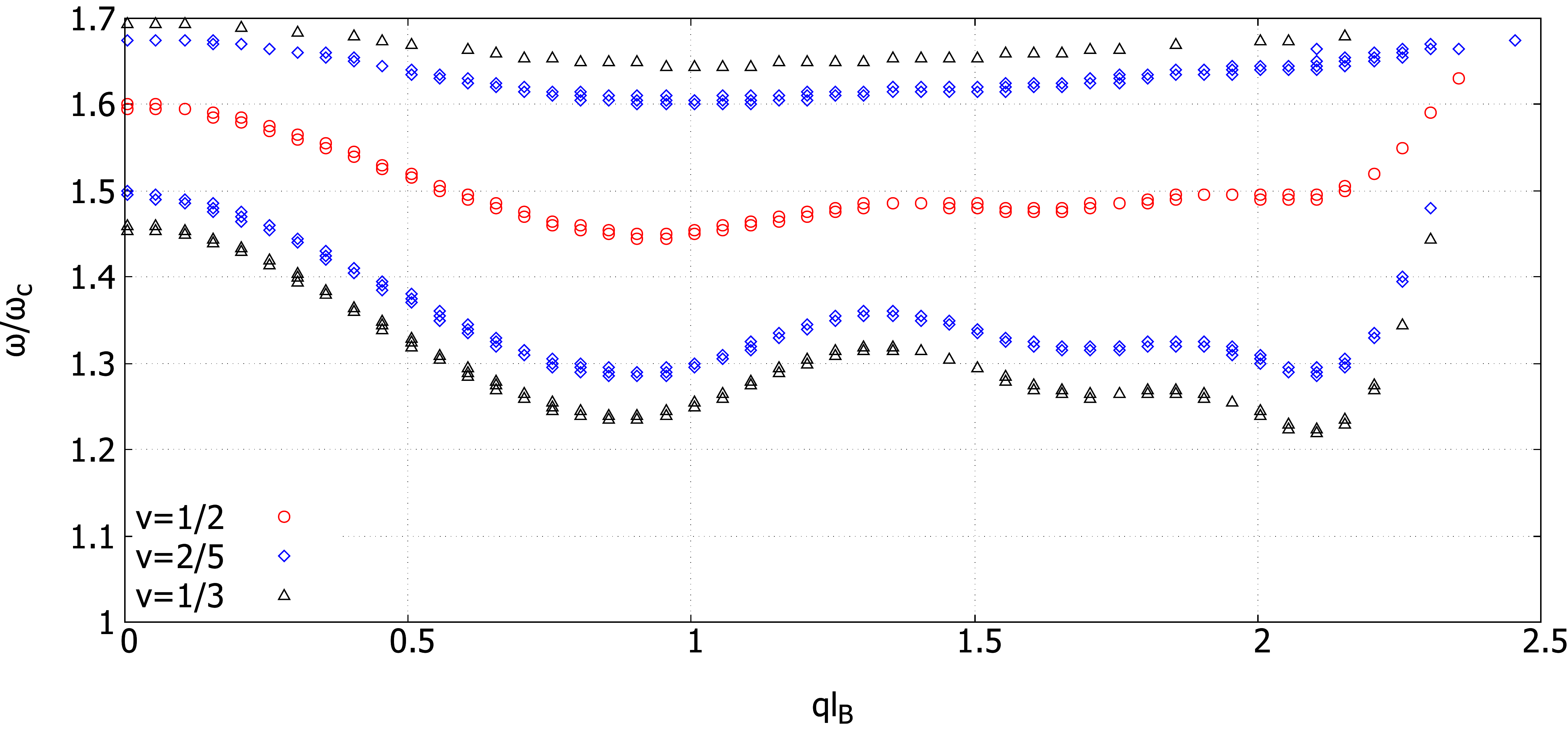}\quad
\caption{Dispersion curves for $\nu =1/2$ (1,1,2,1,0,0), $\nu=1/3$ (2,1,2,1,0,0) and $\nu=2/5$ (2,0,3,1,0,0). For this plot $(C, F, N)$ is the same for all three fractions while $(k,m,n)$ varies. }
\label{fig:fixedCNF}
\end{figure*}

For the spectra in Fig.~\ref{fig:fixedkmn} we consider the simplest flux attachment, $(k,m,n)$, that is consistent with all three of these states. 
 We found that for this case different CF-LLs are occupied, translating to different order parameter combinations, $(C, F, N)$. 
 The $\nu=1/3$ and the $\nu=2/5$ state have the same spectra and a higher gap, $\omega_g$ than the $\nu=1/2$ states. 
 For the spectra in Fig.~\ref{fig:fixedCNF} we consider CF-LL fillings (equivalently combinations of CSB orders) that are the same for all three FQH states. 
 In this case the flux attachment parameters $(k,m,n)$ differ in each case. 
 We observe that in this case the $\nu=1/2$ state is the most stable followed by the $\nu=2/5$ and 
 then the $\nu=1/3$ state. The results are summarized in Tables~\ref{tab:table4} and \ref{tab:table5}.

\begin{table}
\caption{$\nu = 1/3$, $\nu=1/2$ and $\nu=2/5$ for fixed $(k, m, n)$}
\label{tab:table4}
\begin{tabular}{|c|c|c|c|c|c|c|c|c|c|c|c|c|c|}
\hline
$\nu$ & k & m & n & C & N & F & $\omega_g/\omega_c$ & $q_{rm}^1 l_B$ & $q_{rm}^2 l_B$ & $q_{rm}^3 l_B$ & $\Delta_r^1/\omega_c$ & $\Delta_r^2/\omega_c$ & $\Delta_r^3/\omega_c$ \\ \hline 
1/3 & 1 & 1 & 2 & 1 & 1 & 1 & 1.62 & 0.97 & 1.60 & 2.06 & 1.49 & 1.52 & 1.53  \\
2/5 & 1 & 1 & 2 & 1 & -1 & -1 & 1.62 & 0.97 & 1.60 & 2.06 & 1.49 & 1.52 & 1.53 \\
1/2 & 1 & 1 & 2 & 1 & 0 & 0 & 1.59 & 0.92 & 1.63 & 2.07 & 1.45 & 1.48 & 1.49  \\
\hline
\end{tabular}
\end{table}

\begin{table}
\caption{$\nu = 1/3$, $\nu=1/2$ and $\nu=2/5$ for fixed $(C, F, N)$}
\label{tab:table5}
\begin{tabular}{|c|c|c|c|c|c|c|c|c|c|c|c|c|c|}
\hline
$\nu$ & k & m & n & C & N & F & $\omega_g/\omega_c$ & $q_{rm}^1 l_B$ & $q_{rm}^2 l_B$ & $q_{rm}^3 l_B$ & $\Delta_r^1/\omega_c$ & $\Delta_r^2/\omega_c$ & $\Delta_r^3/\omega_c$ \\ \hline 
1/2 & 1 & 1 & 2 & 1 & 0 & 0 & 1.59 & 0.92 & 1.63 & 2.07 & 1.45 & 1.48 & 1.49   \\
2/5 & 2 & 0 & 3 & 1 & 0 & 0 & 1.49 & 0.91 & 1.71 & 2.11 & 1.29 & 1.32 & 1.29 \\
1/3 & 2 & 1 & 3 & 1 & 0 & 0 & 1.46 & 0.89 & 1.70 & 2.10 & 1.24 & 1.26 & 1.22 \\
\hline
\end{tabular}
\end{table}

  \begin{figure*}[htb]
\includegraphics[width=16cm]{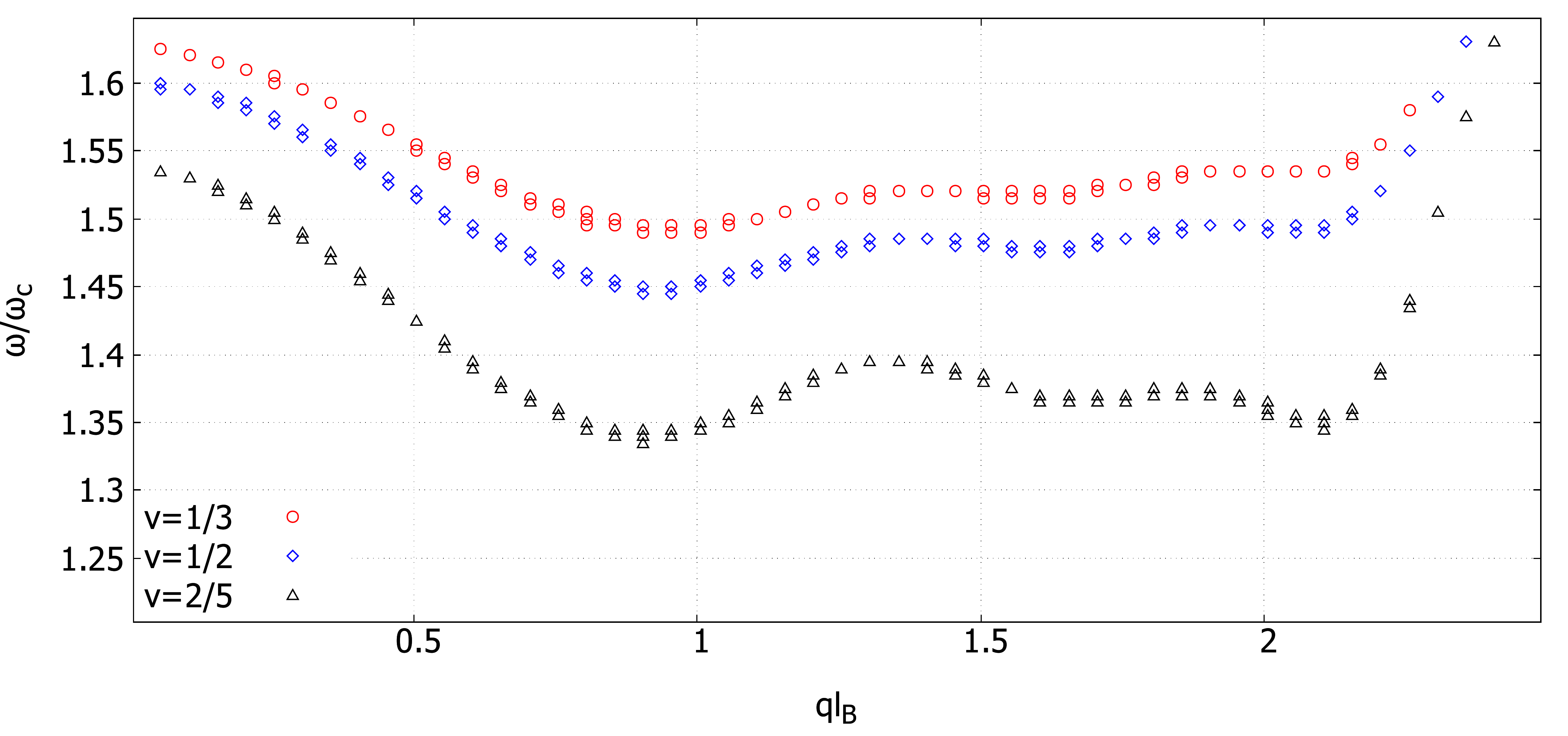}\quad
\caption{Dispersion curves for $\nu =1/3$ (1,1,2,1,1,1), $\nu=2/5$ (1,1,2,1,0,0) and $\nu=2/5$ (1,4,2,1,1,1). }
\label{fig:morenu}
\end{figure*}

\end{widetext}

\section{Discussion and Conclusions}

In this paper we considered the collective excitations for various fractional quantum Hall states in graphene. In doing so we allowed for 
CSB orders which arise naturally in some variational wavefunctions for the FQH states in the zeroth LL, for a variety of flux attachment schemes \cite{Modak2011, Narayanan2018}. 

The interaction induced IQH states at $\nu =0, 1$ have both been suggested as being connected to CSB orders \cite{Herbut2007a, Herbut2007b, Roy2014, Khveshchenko2001, Herbut2006,HJR2009} with good agreement betweeen experimentally measured gaps and theoretical predictions based on a CSB scenario \cite{Roy2014}. The observation of even denominator FQH (EDFQH) states at $\nu= 1/2$ and $\nu=1/4$ \cite{Zibrov2018} has also been suggested as a consequence of CSB orders \cite{Narayanan2018}. 

We primarily studied the $\nu =1/3$ FQH state, considering various flux attachment schemes parameterized by $\mathcal{K}$. We considered the collective excitations for two classes of variational states - the T{\"{o}}ke-Jain sequence \cite{Csaba2007}, parameterized by ($\nu^*$,$k$) and the MMS sequence \cite{Modak2011, Narayanan2018}, parameterized by ($k,m,n$). We found that the MMS states displayed a larger $q \rightarrow 0$ gap, $\omega_g$, and a larger magnetoroton gap than the T{\"{o}}ke-Jain states for the variational states we considered. For the MMS states we considered, we found that increasing $k$, $m$ or $n$ generally reduced $\omega_g$ and the magnetoroton gap. 

Within the framework of the Chern-Simons theory that we use to obtain the collective excitation spectrum, the larger gaps we find for MMS states with low values of ($k,m,n$) suggest that these are likely to be the most stable FQH states. However, there is the caveat that as a mean-field like theory, the Chern-Simons approach will almost certainly over-estimate energy gaps and it is unknown whether fluctuations beyond mean field theory will differ between  T{\"{o}}ke-Jain and MMS states, although we see no a priori reason why they should be significantly different in the two cases. 

We compared the excitation spectra for MMS states for several different fractions, $\nu=1/3$, $2/5$ and $1/2$. 
The true nature of the ground states for these fractions are not currently known \cite{Zibrov2018,Narayanan2018}.
Hence we considered a variety of variational ground states of the MMS type that give rise to incompressible 
states at these fractions.  We compared excitations for states with the same flux attachment scheme (fixed $(k, m, n)$), but different
composite fermion LL fillings (corresponding to differing order parameters  $(C, F, N)$).  We also compared excitations for states
with the same order parameters $(C, F, N)$ but different flux attachment schemes.  The ordering of the states in terms of which had the largest
gap $\omega_g$ is different in the two cases.  Experimental observations (Fig. S14a in Ref.~\cite{Zibrov2018}) show the 
$\nu = 1/3$ and $\nu = 1/2$ states persisting to a temperature of $T = 2.0$ K, while the $\nu=2/5$ state is no longer present at that temperature.
Our calculation of the collective mode spectra do not take into account temperature dependence of the order parameters $C$, $F$ and $N$ and so 
it is not possible to make a direct comparison between our results and experiment.  However, we do give an example of states which lead to the 
same ordering in $\omega_g$ as the stability of the experimental states in Fig.~\ref{fig:morenu}.


In addition to the energy scales associated with the collective excitations, the position of the magnetoroton minimum is also a quantity of interest. Previous work \cite{Simon-Halperin, Balram2017, Golkar2016, Wang2017} has elaborated on the position of the magnetoroton minimum for fractions in the Jain sequence,  $\nu = s/(2s + 1)$, and it was found that for low values of $s$ the minimum was located around $ql_B \sim 1$. Ref. \cite{Balram2017} calculated the positons of minima for $n=0,1$ LLs in graphene.  This is in agreement with our observations in Sec. V, although we find that for both T{\"{o}}ke-Jain and MMS states, there may be secondary and tertiary  minima for $ql_B \sim 1.7$ and $ql_B \sim 2$ respectively. \par

In our calculations here we have shed some light on collective excitations for certain classes of FQH states in graphene. The approach we have taken is restricted to order parameters that can be written in terms of just spin and valley degrees of freedom, such as CDW, N{\'e}el AFM and ferromagnetic order, For more general orders that may require an eight component Dirac fermion description, such as in-plane antiferromagnetism \cite{Roy2014} or partially sublattice polarized (PSP) order \cite{Zibrov2018}, we are unable to calculate the collective excitation spectra. This is because for those more general orders, the order parameter leads to a problem which is mathematically equivalent to one in which one is trying to calculate the collective excitations in a system where there is tunnelling between two separate FQH systems. We are not aware of any succesful attempts to use Chern-Simons approaches to calculate collective excitation spectra in FQH systems with tunnelling between layers.\par
In conclusion, we eagerly look forward to experimental measurements of collective excitation spectra of FQH states in graphene as these will place additional constraints on theoretical approaches and refine our knowledge of the broken symmetries in the zeroth LL in graphene. 

\section{Acknowledgments}
S. N. and M. K. acknowledge support from NSERC. The authors thank A. Mokhtari-Jazi for useful discussions and J. M. Lucero and R. Sarkar for help with the numerical calculations and Compute Canada for access to their resources.

\begin{appendix}
\begin{widetext}

	\section{Calculation of Polarization Tensor and Denominator matrix}
	\label{app:calc}
In this appendix we give more details on the calculation of the polarization tensor that we use to find collective modes. 
We begin with Eq. $\eqref{eqn:28}$ for the electromagnetic response tensor
\begin{equation}
\bf{K} = \bf{\Pi - \Pi (\Pi + C + \mathcal{G})^{-1} \Pi},
\end{equation}
where ${\bf \Pi}$ is the polarization tensor, ${ \bf C}$ is the Chern-Simons tensor and ${\bf \mathcal{G}}$ is a gauge fixing term added to the Lagrangian to make the inverse term finite.
As mentioned in Sec. IV, due to the transverse nature of the polarization tensor it can be decomposed in terms of scalars, $\Pi^0$, $\Pi^1$ and $\Pi^2$, as shown in Eq. $\eqref{eqn:{17-22}}$. Here we are interested in the retarded component of the polarization tensor $(\Pi^i)^R$ and for simplicity we have dropped the $R$. 

The polarization tensor in the absence of order parameters (we discuss the expression in the presence of symmetry breaking mass terms in Appendix~\ref{app:order}) can be expressed as \cite{Frassdorf2018}: 

\begin{equation}~\label{eqn:A2}
 \Pi_{\alpha}^{\mu \nu}(\omega, {\bf{q}}) = \frac{1}{32 \pi^2 l_{\alpha}^4}  \sum_{n,n'=0}^{\infty} \sum_{\lambda, \lambda^{'}=\pm 1} \frac{\mathcal{F}^{\lambda \lambda^{'}}_{{n,n'}}(T, \mu_{\alpha})}{\omega - \lambda\sqrt{n}\omega_{c}^{\alpha} + \lambda^{'}\sqrt{n'}\omega_{c}^{\alpha}}  \int_{\Delta {\bf{r}}} e^{-i{\bf{q}}\cdot\Delta{\bf{r}}}e^{-\Delta{\bf{r}}^2/l_{\alpha}^2} {\rm{Tr}}\left[\sigma_{\alpha}^{\mu}M^{\alpha}_{n}(\lambda \Delta {\bf{r}})\sigma_{\alpha}^{\nu}M^{\alpha}_{n'}(-\lambda^{'}\Delta {\bf{r'}})\right] ,
\end{equation}
 with 
\begin{equation}
\mathcal{F}^{\lambda \lambda^{'}}_{n,n'}(T, \mu_{\alpha}) ={\rm tanh}\left(\frac{\lambda^{'}\sqrt{n'}\omega_{c}^{\alpha} - \mu_{\alpha}}{2T}\right) - {\rm tanh}\left(\frac{\lambda \sqrt{n}\omega_{c}^{\alpha} - \mu_{\alpha}}{2T}\right).
\end{equation}
Here, $T$ is the temperature, $\mu_{\alpha}$ is the chemical potential of species $\alpha$, $\lambda(\lambda^{'}) = \pm 1$ refers to the conduction ($+1$) band or valence ($-1$) band, $\sigma^{\mu}_{\alpha} = (\sigma_0, \kappa^{\alpha} v_F \sigma_1, \kappa^{\alpha} v_F \sigma_2)$, $\omega_c = \sqrt{2}\frac{v_F}{l_\alpha}$ and 
\begin{equation}
M^{\alpha}_{n}(\lambda \Delta {\bf{r}}) = \mathcal{P}_{+} L_n^{0}\left(\frac{\Delta{\bf{r}}^2}{2 l_{\alpha}}\right) + \mathcal{P}_{-} L_{n-1}^{0}\left(\frac{\Delta{\bf{r}}^2}{2 l_{\alpha}}\right) + i \frac{\lambda \kappa}{\sqrt{2} l_{\alpha}} \frac{{\bf{\sigma}} \cdot \Delta {\bf{r}}}{\sqrt{n}}L^{1}_{n-1}\left(\frac{\Delta{\bf{r}}^2}{2 l_{\alpha}}\right),
\end{equation} 
where $L_n^{k}$ is a generalized Laguerre polynomial and $\mathcal{P}_{\pm}$ are projection operators on the sublattice space defined as:
\begin{equation}
\mathcal{P}_{\pm} = \frac{1}{2} \left[ \sigma_0 \pm {\rm sign}(e \mathcal{B}^{{\rm eff}}_{\alpha})\sigma_3 \right].
\end{equation}

 The scalars $\Pi^0$, $\Pi^1$ and $\Pi^2$ are given by:
\begin{equation}{\label{eqn:A6}}
 \Pi_{\alpha}^{0}(\omega, \vec{q}) = \frac{-1}{32 \pi^2 l_{\alpha}^4 \vec{q}^2} \sum_{n,n'=0}^{\infty} \sum_{\lambda, \lambda^{'} = \pm 1} \frac{\mathcal{F}^{\lambda \lambda^{'}}_{n,n'}(T, \mu_{\alpha})}{\omega - \lambda\sqrt{n}\omega_{c}^{\alpha} + \lambda^{'}\sqrt{n'}\omega_{c}^{\alpha}}   \left[I^{0}_{n-1,n'}(\mathcal{Q}_{\alpha}) + I^{0}_{n,n'-1}(\mathcal{Q}_{\alpha})  + \frac{2\lambda \lambda^{'}}{\sqrt{n n'}}I^{1}_{n-1,n'-1}(\mathcal{Q}_{\alpha})\right],
 \end{equation}
 \begin{equation}
 \Pi_{\alpha}^{1}(\omega, \vec{q}) = -\frac{v_F^2}{32 \pi^2 l_{\alpha}^4 \omega} \sum_{n,n'}^{\infty} \sum_{\lambda, \lambda^{`} = \pm 1} \frac{\mathcal{F}^{\lambda \lambda'}_{n,n'}(T, \mu_{\alpha})}{\omega - \lambda\sqrt{n}\omega_{c}^{\alpha} + \lambda^{'}\sqrt{n'}\omega_{c}^{\alpha}}  \left[I^{0}_{n-1,n'}(\mathcal{Q}_{\alpha}) - I^{0}_{n,n'-1}(\mathcal{Q}_{\alpha})\right],
 \end{equation}
 \begin{equation}{\label{eqn:A8}}
\Pi_{\alpha}^{2}(\omega, \vec{q}) = \frac{v_F^2}{32 \pi^2 l_{\alpha}^2 } \sum_{n,n'=0}^{\infty} \sum_{\lambda, \lambda^{'}= \pm 1} \frac{\mathcal{F}^{\lambda \lambda '}_{n,n'}(T, \mu_{\alpha})}{\omega - \lambda\sqrt{n}\omega_{c}^{\alpha} + \lambda^{'}\sqrt{n'}\omega_{c}^{\alpha}} \left[\frac{2\lambda \lambda^{'}}{\sqrt{n n'}}\partial^{2}_{\mathcal{Q}_{\alpha}}\tilde{I}^{1}_{n-1,n'-1}(\mathcal{Q}_{\alpha})\right].
\end{equation}
The terms $I_{n,n'}^{k}$ and $\tilde{I}_{n,n'}^{k}$ are given by \cite{Frassdorf2018}
\begin{eqnarray}
I_{n,n'}^{k}(\mathcal{Q_{\alpha}}) &=& 2\pi  \ell_{\alpha}^{2} \mathcal{Q}_{\alpha}^{(n_> - n_<)} e^{-\mathcal{Q}_{\alpha}}\frac{(n_< + k)!}{n_> !} L_{n_<}^{(n_> - n_<)}(\mathcal{Q}_{\alpha}) L_{(n_< + k)}^{(n_> - n_<)}(\mathcal{Q}_{\alpha}),\\ \nonumber
\tilde{I}^1_{n-1,n'-1}(\mathcal{Q_{\alpha}}) &=& \sum_{m=0}^{n-1} \sum_{m'=0}^{n'-1} I^{0}_{m,m'}(\mathcal{Q}_{\alpha}).
\end{eqnarray}
where $\mathcal{Q}_{\alpha} = q^2 l_B^2/2$, $n_< = {\rm min}\{n,n'\}$ and $n_> ={\rm max}\{n,n'\}$. Clearly both $\tilde{I}^k_{n-1,n'-1}(\mathcal{Q_{\alpha}})$ and $I_{n,n'}^{k}(\mathcal{Q_{\alpha}})$ are symmetric in the indices $n$ and $n'$.

\section{Sums of Laguerre Polynomials}
To evaluate the scalars, $\Pi^0$, $\Pi^1$ and $\Pi^2$ we have to evaluate sums of the form

\begin{equation}
S^{b} = \sum_{n,n'=0}^{\infty} \sum_{\lambda, \lambda^{'}=\pm 1} \frac{{\rm tanh}\left( \frac{\lambda' \sqrt{n'}\omega_{c}^{\alpha} - \mu_{\alpha} }{2T}\right) - {\rm tanh}\left (\frac{\lambda\sqrt{n}\omega_{c}^{\alpha} - \mu_{\alpha} }{2T}\right)}{(\omega \pm i0) - \lambda \sqrt{n}\omega_{c}^{\alpha} + \lambda' \sqrt{n'}\omega_{c}^{\alpha}} {\huge{T^{k,b}_{n,n'}}},
\end{equation}
and of the form 
\begin{equation}
S_1^{b} = \sum_{n,n'=0}^{\infty} \sum_{\lambda, \lambda^{'}=\pm 1} \frac{2\lambda \lambda'}{\sqrt{n n'}} \frac{{\rm tanh}\left( \frac{\lambda' \sqrt{n'}\omega_{c}^{\alpha} - \mu_{\alpha} }{2T}\right) - {\rm tanh}\left (\frac{\lambda\sqrt{n}\omega_{c}^{\alpha} - \mu_{\alpha} }{2T}\right)}{(\omega \pm i0) - \lambda \sqrt{n}\omega_{c}^{\alpha} + \lambda' \sqrt{n'}\omega_{c}^{\alpha}} {\huge{T^{k,b}_{n,n'}}} 
\end{equation}
where $b = 0,1,2$. The $T^{k,b}_{n,n'}$ are terms involving $I^{k}_{n,n'}$, $\tilde{I}_{n,n'}^{k}$ present in $\Pi^b$ .
We have,
\begin{equation}
 n_F(\xi) = 1/2 \left[1 - {\rm tanh}\left(\frac{\beta}{2}\xi\right)\right] \Rightarrow {\rm tanh}\left(\frac{\beta}{2}\xi \right) =1 - 2n_{F}(\xi), \nonumber
 \end{equation}
where $n_F(\xi)$ is the Fermi-Dirac distribution and in the limit  $T \rightarrow 0$, $n_F$ becomes a step function 
and we can write the sum as
\begin{equation}~\label{eqn:B2}
S^b =2 \sum_{n,n'=0}^{\infty} \sum_{\lambda, \lambda^{`}=\pm 1} \frac{\Theta(\lambda \sqrt{n}\omega_{c}^{\alpha} - \mu_{\alpha}) - \Theta (\lambda^{`}\sqrt{n'}\omega_{c}^{\alpha} - \mu_{\alpha})}{(\omega \pm i0) - \lambda \sqrt{n}\omega_{c}^{\alpha} + \lambda' \sqrt{n'}\omega_{c}^{\alpha}} {\huge{T^{k,b}_{n,n'}}}.
\end{equation}
\\
Now, depending on whether $\lambda, \lambda'$ is $\pm$ we have 4 different possible terms. The chemical potential can be either positive or negative. We begin with the case for positive chemical potential.  \\
\subsection{Positive Chemical Potential}
Evaluating $S^b$ for $\mu_{\alpha} > 0 $ gives

\begin{eqnarray}
S^b&=& 2 \sum_{n'=0}^{\beta}\sum_{n=\beta + 1}^{\infty}  \frac{1}{(\omega \pm i0) - \sqrt{n}\omega_{c}^{\alpha}
 + \sqrt{n'}\omega_{c}^{\alpha}}T^{k,b}_{n,n'} - 2 \sum_{n=0}^{\beta}\sum_{n'=\beta + 1}^{\infty}  \frac{1}{(\omega \pm i0) 
 - \sqrt{n}\omega_{c}^{\alpha} + \sqrt{n'}\omega_{c}^{\alpha}}T^{k,b}_{n,n'}\\ \nonumber
 & &  -2 \sum_{n=0}^{\infty}\sum_{n'=\beta + 1}^{\infty}
   \frac{1}{(\omega \pm i0) + \sqrt{n}\omega_{c}^{\alpha} + \sqrt{n'}\omega_{c}^{\alpha}} T^{k,b}_{n,n'}
    +2  \sum_{n'=0}^{\beta}\sum_{n=\beta + 1}^{\infty}  \frac{1}{(\omega \pm i0)
     - \sqrt{n}\omega_{c}^{\alpha} - \sqrt{n'}\omega_{c}^{\alpha}}T^{k,b}_{n,n'},
\end{eqnarray}
with $\beta = \lfloor{\frac{\mu_{\alpha}^2}{(\omega_{c}^{\alpha})^2}}\rfloor$.

\subsection{Negative Chemical Potential}
Evaluating for $S^b$ for $\mu_{\alpha} < 0$ gives
\begin{eqnarray}
S^b &=& 2 \sum_{n=0}^{\infty}  \sum_{n'=\beta + 1}^{\infty}  \frac{1}{(\omega \pm i0) - \sqrt{n}\omega_{c}^{\alpha} - \sqrt{n'}\omega_{c}^{\alpha}}T^{k,b}_{n,n'} - 2 \sum_{n=0}^{\beta}  \sum_{n'=0}^{\infty}  \frac{1}{(\omega \pm i0) + \sqrt{n}\omega_{c}^{\alpha} + \sqrt{n'}\omega_{c}^{\alpha}}T^{k,b}_{n,n'} \\ \nonumber
& & +2 \sum_{n=0}^{\beta}  \sum_{n'=\beta + 1}^{\infty}  \frac{1}{(\omega \pm i0) + \sqrt{n}\omega_{c}^{\alpha} - \sqrt{n'}\omega_{c}^{\alpha}}T^{k,b}_{n,n'}  - 2 \sum_{n'=0}^{\beta}  \sum_{n=\beta + 1}^{\infty}  \frac{1}{(\omega \pm i0) + \sqrt{n}\omega_{c}^{\alpha} - \sqrt{n'}\omega_{c}^{\alpha}}T^{k,b}_{n,n'}.
\end{eqnarray}

Since the terms $I_{n,n'}^{k}$ and $\tilde{I}_{n,n'}^{k}$ are  symmetric in $n$ and $n'$, the expressions for the sums are exactly the same for $\mu_{\alpha} > 0$ and $\mu_{\alpha} < 0$ if we exchange $n$ and $n'$ which are just dummy variables. We can hence simplify $S^b$ further by combining the terms carefully to get the final expressions as:

\begin{equation}
S^b =\left[ 2\sum_{n' =0}^{\beta} \sum_{n=\beta+1}^{\infty} \frac{2\sqrt{n}\omega_{c}^{\alpha}\left[ (\omega)^2 - (n-n')(\omega_{c}^{\alpha})^2\right]}{\omega^4 -\omega^2(\omega_{c}^{\alpha})^2(n+n')+ (n - n')^2(\omega_{c}^{\alpha})^4} +  2\sum_{n'=\beta + 1}^{\infty} \sum_{n=\beta+1}^{\infty} \frac{2 \omega_{c}^{\alpha} (\sqrt{n} + \sqrt{n'})}{\omega^2 - (\omega_{c}^{\alpha})^2 (\sqrt{n} + \sqrt{n'})^2}\right] T^{k,b}_{n,n'}.
\end{equation}

Our expression for  $\Pi^1$ [Eq. $\eqref{eqn:A6}]$ then becomes:
 \begin{equation}~\label{eqn:D4}
 \Pi^{1}_{\alpha} =  -\frac{v_F^2}{32 \pi^2 l_{\alpha}^4 \omega} S^1 .
 \end{equation}

 In $\Pi^0$ and $\Pi^2$ [Eq. $\eqref{eqn:A6}$] and [Eq. $\eqref{eqn:A8}$] respectively, we have a multiplicative factor of $ \frac{2\lambda \lambda^{'}}{\sqrt{n n'}}$  .  Taking into account the multiplicative factor we get the sum, $S_1^b$, over $\lambda, \lambda'$ and $n, n'$ to be: 

\begin{equation}
 S_1^b = \left[- 4 \sum_{n=0}^{\infty} \sum_{n'=\beta +1}^{\infty}\frac{1}{\sqrt{n n'}} \frac{ \omega_{c}^{\alpha}(\sqrt{n} + \sqrt{n'})}{\omega^2 - (\sqrt{n} + \sqrt{n'})^2 (\omega_{c}^{\alpha})^2} + 
   4 \sum_{n'=0}^{\beta} \sum_{n=\beta +1}^{\infty}\frac{1}{\sqrt{n n'}}  \frac{ \omega_{c}^{\alpha}(\sqrt{n} - \sqrt{n'})}{\omega^2 - (\sqrt{n} - \sqrt{n'})^2 (\omega_{c}^{\alpha})^2} \right]T^{k,b}_{n,n'}.
   \end{equation}

 With this we can write down the expression for $\Pi^0$ as,
\begin{equation}	~\label{eqn:D6}
 \Pi^0_{\alpha} =  - \frac{1}{32\pi^2 l_{\alpha}^{4} q^2} [ S^0  + S_1^0 ].
 \end{equation}

 Similarly, $\Pi_2$ is given by,
 \begin{equation}~\label{eqn:D7}
 \Pi^2_{\alpha} = \frac{v_F^2}{32 \pi^2 l_{\alpha}^2} (S_1^2 )
 \end{equation}

\section{Effect of order parameters}
\label{app:order}

In the presence of order parameters the energy eigenvalues change, $\epsilon \sim \sqrt{n} \rightarrow \epsilon \sim \sqrt{n + \Delta^2} $ where $\Delta$ is the gap arising from linear combinations of $m_{\alpha}$ and $f_{\alpha}$. The arguments from Appendix B. i.e. without the order parameters, still hold, and hence the expressions for the sums over $\lambda, \lambda'$ and $n, n'$'s for $S$, as defined in the previous section, becomes,
\begin{eqnarray}~\label{eqn:E1}
S^b &=& 2\sum_{n' =0}^{\beta} \sum_{n=\beta+1}^{\infty} \frac{2\sqrt{n+ m_{\alpha}^{2}}\omega_{c}^{\alpha}\left[ (\omega)^2 - (n-n')(\omega_{c}^{\alpha})^2\right]}{\omega^4 -\omega^2(\omega_{c}^{\alpha})^2(n+n' +2 m_{\alpha}^{2}) + (n - n')^2(\omega_{c}^{\alpha})^4}T^{k,b}_{n,n'} \\ \nonumber
 & & +2\sum_{n'=\beta + 1}^{\infty} \sum_{n=\beta+1}^{\infty} \frac{2 \omega_{c}^{\alpha} (\sqrt{n+ m_{\alpha}^{2}} + \sqrt{n'+ m_{\alpha}^{2}})}{\omega^2 - (\omega_{c}^{\alpha})^2 (\sqrt{n+ m_{\alpha}^{2}} + \sqrt{n'+ m_{\alpha}^{2}})^2}T^{k,b}_{n,n'},
 \end{eqnarray}
and
\begin{eqnarray}~\label{eqn:E2}
 S_1^b &=& - 4 \sum_{n=0}^{\infty} \sum_{n'=\beta +1}^{\infty} \frac{ \omega_{c}^{\alpha}(\sqrt{n+ m_{\alpha}^{2}} + \sqrt{n'+ m_{\alpha}^{2}})}{\omega^2 - (\sqrt{n+ m_{\alpha}^{2}} + \sqrt{n'+ m_{\alpha}^{2}})^2 (\omega_{c}^{\alpha})^2}\frac{1}{\sqrt{n n'}}T^{k,b}_{n,n'} \\ \nonumber 
 & & + 4 \sum_{n'=0}^{\beta} \sum_{n=\beta +1}^{\infty} \frac{ \omega_{c}^{\alpha}(\sqrt{n+ m_{\alpha}^{2}} - \sqrt{n'+ m_{\alpha}^{2}})}{\omega^2 - (\sqrt{n+ m_{\alpha}^{2}} - \sqrt{n'+ m_{\alpha}^{2}})^2 (\omega_{c}^{\alpha})^2}\frac{1}{\sqrt{n n'}}  T^{k,b}_{n,n'}.
 \end{eqnarray}

\end{widetext}
\end{appendix}

\bibliographystyle{apsrev4-1}
\bibliography{arxiv}

\end{document}